\documentclass[12pt]{article}
\usepackage[margin=2cm]{geometry}
\usepackage{comment}
\usepackage{amsmath,amssymb,extarrows,mathtools,graphicx,subfigure,setspace}
\usepackage{cite}
\usepackage{slashed}
\usepackage{color}
\makeatother

\def\R{{\rm I\hspace{-.15em}R}}

\def\R{{\rm I\hspace{-.15em}R}}

\newcommand{\bei}{\begin{itemize}}
\newcommand{\eni}{\end{itemize}}

\numberwithin{equation}{section}
\begin{document}
\onehalfspacing
\vfill
\begin{titlepage}
\vspace{10mm}
\begin{center}
{\Large {\bf Quasinormal modes in de Sitter space: Plane wave method }\\
}

\vspace*{15mm}
\vspace*{1mm}
{M. Reza Tanhayi}

 \vspace*{1cm}

{\it  Department of Physics, Islamic Azad University, Central Tehran Branch, Tehran,
 Iran\\
}
 \vspace*{0.5cm}
{E-mail: {\tt m$_{-}$tanhayi@iauctb.ac.ir}}%

\vspace*{1cm}
\end{center}

\begin{abstract}
Recently, in the context of dS/CFT correspondence, quasinormal modes have been put forward to address certain features of this conjecture. In particular, it is argued that the dual states of quasi-normal modes are in fact the states of CFT$_3$ which are created by operator insertions. For a scalar field in $dS_4$, quasi-normal modes which are singular on the past horizon of the south pole and decay exponentially towards the future have been considered in \cite{Ng:2012xp, Jafferis:2013qia}, these modes lie in two complex highest-weight representation of the dS$_4$ isometry group. In this work, we present a simple group representation analysis of these modes so that the de Sitter invariance is obviously manifest. By making use of the so-called plane wave method, we will show that the quasi-normal modes correspond to one class of the unitary irreducible representation of the de Sitter group. This consideration could be generalized straightforwardly for higher-spin fields and higher dimensions, in particular, we will study the quasinormal modes for gauge and spinor fields, and, in the case of a scalar field, the generalization to higher dimensions is also obtained. 

\end{abstract}
\end{titlepage}

\section{Introduction}

The study of quasinormal modes (QNMs) is of great importance and has a rich history (see \cite{Kokkotas:1999bd, Berti:2009kk, Konoplya:2011qq} for a comprehensive review). In black hole physics, QNMs are defined by a set of damped oscillations produced due to a perturbation around the black holes. They are in fact the single-frequency oscillations which satisfying ingoing and outgoing boundary conditions, respectively at the event horizon and spatial infinity of the black hole \cite{Price:1991xe}. The QNMs completely are determined by the black hole's parameters and, in principle, they can carry out the information about the black holes; e.g., they play a significant role in studying the black hole's area quantization \cite{Hod:1998vk, Corichi:2002ty, Dreyer:2002vy}. On the other hand in the context of the proposed correspondence between the classical gravity in anti-de Sitter space and conformal field theory living on the boundary, QNMs play a crucial role in studying the dynamical time scale of thermal equilibrium or thermalization in a strongly coupled conformal field theory \cite{Horowitz:1999jd}. But, unlike the AdS/CFT, there is no such satisfactory dictionary between the bulk theory with a positive cosmological constant and its dual conformal field theory, and more investigation is actually needed to better understand the de Sitter space. Therefore, achieving a proper dS/CFT correspondence, besides the astrophysical motivations of course, makes de Sitter important \cite{Strominger:2001pn, Strominger:2001gp}. Especially, the QNMs in this background have been considered more and it is argued that the QNMs spectrum is related to the thermal excitations of the boundary conformal field theory at past and future infinities of de Sitter space \cite{Abdalla:2002hg}.    

Quasinormal modes have been studied in several methods in a de Sitter background \cite{Natario:2004jd, LopezOrtega:2006my, Brady:1999wd, Choudhury:2011at}, but, in a manifestly de Sitter invariant way, these modes have been found in Ref.\cite{Ng:2012xp} as the modes of a scalar field which lie in two complex highest-weight representations of the isometry group of de Sitter space; these modes are singular on the past horizon of the south pole and decay exponentially towards the future. In four dimensional de Sitter global coordinates $x=(t,\theta_1,\theta_2,\theta_3)$ with line element (in units the Hubble constant $H$ is set to be one) 
\begin{equation}\label{global metric}
ds^2=-dt^2+\cosh^2t \,\,\,d\Omega^2_3,\hspace{3mm} -\infty<t<\infty,
\end{equation} 
the north and south poles are denoting by 
\begin{equation}
\Omega_{SP}\sim \theta_1=0,\hspace{3mm} \Omega_{NP}\sim \theta_1=\pi,
\end{equation}
in which $\Omega=(\theta_1,\theta_2,\theta_3)$ are the angle coordinates on $S^3$ with
\begin{equation}
\nonumber
0\leq \theta_1,\,\theta_2<\pi,\hspace*{3mm}0\leq \theta_3<2\pi.
\end{equation}
In de Sitter global coordinates, for a light massive scalar field of mass $m^2 <9/4$, the quasinormal modes are defined by 
\begin{equation}\label{QNM}
\phi_{QN}^{\pm}(x)={\cal N_\pm}\frac{1}{(\sinh t-i\epsilon+\cosh t\,\,\cos\theta_1\Big)^{h_\pm}},
\end{equation}
where ${\cal N_\pm}=\frac{\Gamma(\mp\mu)\Gamma(h_\pm)}{4\pi^{\frac{5}{2}}}(1-e^{\mp 2\pi i\mu})$ and 
\begin{equation}\label{h}
h_\pm=\frac{3}{2}\pm\mu,\hspace{6mm}\mu=\sqrt{\frac{9}{4}-m^2}.
\end{equation}
In order the de Sitter invariance to be manifest, the invariant parameters, say as the Euclidean Green function and the geodesic distance are used in \cite{Ng:2012xp}, which are recalled briefly in the next section. 

By the extension of the Fourier-Helgason transformation in de Sitter space, Bros, Moschella and Gazeau in Refs. \cite{Bros:1994dn, Bros:1995js}, have developed an axiomatic field theory in de Sitter space which is based on analyticity in the complexified Riemannian manifold. They found coordinate-independent plane waves in de Sitter space with the following properties.\\
The plane waves give rise to the thermal interpretation for the Bunch-Davies or Euclidean vacuum in de Sitter space. These modes have no antipodal singularities that also satisfy the zero-curvature limit; namely, in this limit, they coincide with the usual plane wave in Minkowski space. The plane waves are given by
\begin{equation}\label{plane}
\Phi^\pm(X,\xi)=\Big(\theta(X\cdot\xi)+\theta(-X\cdot\xi)e^{\pm 2i\pi\sigma_\pm}\Big)(X\cdot\xi)^{\sigma_\pm},
\end{equation} 
 where $\theta$ is the step function and $\sigma_\pm$ (that hereafter we write it $\sigma$) is the homogeneity degree which is fixed by the group theory. The null five-vector $\xi$ is introduced, which plays the role of the $4$-momentum in flat space in order to label the plane waves in de Sitter space. 
 
In this paper, we show that the quasinormal modes (\ref{QNM}) can indeed be obtained via the plane wave approach and these modes correspond to one class of the unitary irreducible representation (UIR) of the de Sitter group, namely, the complementary series. Since the group theoretical approach is utilized, the de Sitter invariance would be manifest; moreover, this method could potentially generalize to higher spins and dimensions.

This paper is organized as follows: In Sec. $2$, we recall the quasinormal modes in dS$_4$ from \cite{Ng:2012xp, Jafferis:2013qia}. In Sec. $3$, through the plane wave formalism, first we study the scalar QNMs and then generalize those modes to $d$ dimensions in Sec. 4. Higher-spin considerations, namely, the gauge and spinor fields, are also done in Sec. $4$. Section $5$ is devoted to a brief conclusion. Finally, some details on plane waves and de Sitter group are given in the Appendix.

\section{Quasinormal Modes in de Sitter Space}

de Sitter space is a maximally symmetric space with ten Killing vectors, and its isometry group is ten-parameter $SO_0(1,4)$. The de Sitter geometry can be thought of as a timelike hyperboloid in the embedding space, namely, the $(4+1)$-dimensional Minkowski space, and the $SO_0(1,4)$ acts manifestly on the embedding coordinates:
\begin{equation}
M_H=\{X\in\R^5,\eta_{\alpha\beta}X^\alpha X^\beta=-X_0^2+X_1^2+X_2^2+X_3^2+X_4^2=1\,(=H^{-2})\}.
\end{equation}
The corresponding metric reads as
\begin{eqnarray}
&ds^2=\eta_{\alpha\beta}dX^\alpha dX^\beta|_{Const.}=g^{dS}_{\mu\nu}dx^\mu dx^\nu\\
&\alpha,\beta=0,1,..., 4,\hspace*{3mm}\mu,\nu=0,1,2,3. \nonumber
\end{eqnarray}
Different coordinates can be obtained via the different parametrization of the constraint equation \cite{Spradlin:2001pw}. For our purpose, with the metric (\ref{global metric}), the Klein-Gordon equation reads as:
\begin{equation}\label{KG}
(\Box-m^2)\phi(x)=0,
\end{equation}
where $\Box=\frac{1}{\sqrt{-g}}\partial_\mu\Big(\sqrt{-g}g^{\mu\nu}\partial_\nu\Big)$. Being the maximally symmetric space allows one to write $\phi(x)=\chi_L(t)Y_{Lj}(\Omega)$, where the $Y_{Lj}$'s are spherical harmonics on the 3-sphere.
The time evolution equation for $\chi(t)$ then reads as
\begin{equation}
\ddot{\chi}(t)+3\tanh^2t \,\,\dot{\chi}(t)+\Big(m^2+\frac{L(L+2)}{\cosh^2 t}\Big)\chi(t)=0,
\end{equation}
and the future and past behavior of the above equation becomes
\begin{equation}
\ddot{\chi}(t)\pm 3\dot{\chi}(t)+m^2\chi(t)=0,
\end{equation}
where one obtains
\begin{equation}
\begin{array}{l}
\chi(t)\longrightarrow e^{-h_\pm t},\hspace*{3mm} t\rightarrow\infty,\\ \vspace{2mm}
\chi(t)\longrightarrow e^{h_\pm t},\hspace*{4mm} t\rightarrow -\infty
\end{array}
\end{equation}
and $h_\pm$ has been already introduced in (\ref{h}). If one defines 
\begin{equation}
G_\pm(t,\Omega;\,t',\Omega')\equiv G_E(x;x')-e^{i\pi h_\mp}G_E(x;x'_A)
\end{equation}
$x_A$ is the antipodal point\footnote{For any point on de Sitter space (with embedding coordinate $X$), there is a reflected point through the origin of Minkowski space that is named as the antipodal point and one can write $X_A=-X$.} of $x$, then the quasinormal modes are introduced as follows \cite{Jafferis:2013qia}
\begin{equation}
\phi^\pm_{QN}(x)\equiv \lim_{t'\rightarrow\infty}e^{h_\pm t}G_\pm(t,\Omega;t',\Omega'_{NP}),
\end{equation}
where $\Omega'_{NP}$ means that $x'$ is evaluated on the north pole: $\theta_1'=\pi$. Note that $G_E(x,x')$ is the Euclidean Green function \cite{Bousso:2001mw}
\begin{equation}
\nonumber
G_E(x,x')=\frac{1}{8\pi^2}\frac{1}{1-P(x,x')},
\end{equation}
$P(x,x')$ stands for the de Sitter invariant distance which for a given two points on de Sitter hyperboloid, say,  $(t,\Omega)$ and $(t',\Omega')$, is defined by
\begin{equation}
P(x,x')=\cosh t\cosh t'\cos \Theta_3(\Omega,\Omega')-\sinh t\sinh t',
\end{equation}
in which $\Theta_3$ is the geodesic distance
\begin{equation}
\cos\Theta_3(\Omega,\Omega')=\cos\theta_1\cos\theta'_1+\sin\theta_1\sin\theta_1'\Big(\cos\theta_2\cos\theta_2+\sin\theta_2\sin\theta_2'
\cos(\theta_3-\theta_3')\Big).
\end{equation}
These modes are obviously Lorentz invariant. It can be checked that for general light scalar fields namely, $m^2<\frac{9}{4}$, the quasinormal modes in four dimensions are given by (\ref{QNM}). In what follows, we will show that these quasinormal modes can be reproduced through a simple method of plane waves.

\section{quasinormal modes and Plane Wave Method}

In flat space, on-shell $4$-momentum is used to label the free field wave equations which are given by the plane waves; however, having noticed that in de Sitter space there is no global timelike Killing vector, one cannot define a well-defined momentum. Thus, in de Sitter space, more caution is needed in writing a plane wave. The so-called plane wave method was used in Refs.\cite{Bros:1994dn, Bros:1995js} to develop an axiomatic field theory in de Sitter space by constructing the analytic two-point functions. The method goes as follows: The massive scalar field equation in de Sitter space can be viewed as an usual d'Alembert operator in the embedding five-dimensional flat space with a de Sitter constraint
\begin{equation}\label{plkg}
\Big(\eta^{\alpha\beta}\partial^T_\alpha\partial^T_\beta-m^2\Big)\phi(X)=0,
\end{equation}
where $\partial^T$ means that we are interested in the transverse projected derivative on the hyperboloid which is defined by $\partial^T_\alpha=\partial_\alpha-X_\alpha X_\beta\partial^\beta$. The box operator then reads as \cite{Rouhani:2009gh}
\begin{equation}
\Box_H=\eta^{\alpha\beta}\partial^T_\alpha\partial^T_\beta=\partial_\alpha\partial^\alpha-4X_\alpha\partial^\alpha-X_\alpha X_\beta \partial^\alpha\partial^\beta.
\end{equation}
The proper solution to (\ref{plkg}) was found in Ref.\cite{Bros:1994dn}, as a homogeneous function of degree $\sigma$:
\begin{equation}\label{plane wave}
\phi(X)=(X\cdot\xi)^\sigma,
\end{equation}
where the null five-vector $\xi$ can actually play the role of $4$-momentum in flat space. Note that the causal order in $ \R^5$ induces causal order on the de Sitter space. To investigate this, let us define the future (respectively, past) cone as the following subset in $R^5$:
 \begin{equation}\label{forward}
 V_+(=-V_-)=\{X\in\R^5;\,\,X^0>0,\,\,\eta_{\alpha\beta}X^\alpha X^\beta<0\},
 \end{equation}
so that the future (past) light cone is given by 
\begin{equation}
\partial V_+(=-\partial V_-)=\{X\in\R^5;\,\,X^0>0,\,\,\eta_{\alpha\beta}X^\alpha X^\beta=0\}. 
\end{equation} 
For a given point or event, say, $X$, the future (or past) is a set of points $Y\in M_H$ where $Y-X\in V_+$ (or $V_-)$. The future and past shadows of a given event $X\in M_H$  are defined by $\Gamma^+(X)=\{Y\in M_H;\,\,Y\geq X\}$ and $\Gamma^-(X)=\{Y\in M_H;\,\,Y\leq X\}$, respectively. The causal light cone associated with $X \in M_H$ is then defined by $\partial\Gamma^+\cup\partial\Gamma^-$. Thus, if one chooses the null 5-vector, say, as $\xi\in \partial V_+$, it could play the role of $4$-momentum in order to label freely falling wave equations. Namely, the causal order in de Sitter can be understood by the forward asymptotic light cone which is the boundary of the forward cone $V_+$, and the forward-pointing lightlike 5-vector $\xi\in \partial V_+$ can indeed parametrize the plane wave (\ref{plane wave}) in embedding space. This prescription actually mimics the freely moving particles in flat space, where the on-shell 4-momentum $p$ is used to label the plane waves $e^{ip\cdot x}$. The asymptotic cone $\partial V_+$ can then be physically interpreted
as the space of (asymptotic) momenta directions, and in the flat limit ($H\rightarrow 0$) the de Sitter plane wave $(X\cdot \xi)^\sigma$ when it is written in principal series of de Sitter space tends to the flat space plane wave, (more technical details can be found in Refs.\cite{Bros:1994dn, Bros:1995js, Gazeau:2006gq, Bros:1998ik, Moschella:2012dq}). Therefore, after a direct calculation, the massive scalar field equation is obtained as
\begin{equation}\label{mod KG}
\Big(\Box_H+\sigma(\sigma+3)\Big)(X\cdot\xi)^\sigma=0.
\end{equation}
Comparing the mass terms obtained in (\ref{KG}) and (\ref{mod KG}), yields  
\begin{equation}\label{mas}
m^2=-\sigma(\sigma+3),
\end{equation} 
and inserting the above mass in (\ref{h}) leads one to write
\begin{equation}
\sigma_\pm= -\frac{3}{2}\mp \mu,
\end{equation}
comparing with (\ref{h}) reveals that $ \sigma_\pm=-h_\pm$. However, from the group theoretical point of view, $\sigma$ is associated with the eigenvalues of the de Sitter Casimir operators. The de Sitter isometry group has two Casimir operators, and the second-order scalar wave equation in terms of the Casimir operators can be written as
\begin{equation}
\Big({\cal C}_1-\langle{\cal C}_1\rangle\Big)\phi=0,
\end{equation} 
where ${\cal C}_1=-\frac{1}{2}(X_\alpha\partial_\beta-X_\beta\partial_\alpha)^2$ and $\langle{\cal C}_1\rangle$ stands for the eigenvalue of the Casimir operator. One can show that ${\cal C}_1=\Box_H$ and, consequently, 
\begin{equation}\label{eigenvalue}
<{\cal C}_1>= -\sigma(\sigma+3).
\end{equation}
According to the possible values of $\sigma$, three types of UIR of the de Sitter group might be classified: principal, complementary and discrete series.  Relegating the details to the Appendix, here we simply summarize the complementary series in four dimensions and for a scalar field. It is proved that the complementary series in de Sitter space is defined in the following range of $\sigma$:
\begin{equation}
-3<\sigma<0,
\end{equation} 
which in the language of the mass turns to
\begin{equation}
0<m^2<\frac{9}{4},
\end{equation}
where exactly the light mass range of the quasinormal modes considered in Ref.\cite{Jafferis:2013qia} is. 

The mode functions (\ref{plane}) that we have already discussed are not yet normalized. Noting that the mass term in (\ref{mod KG}) is invariant under the change $\sigma\rightarrow -\sigma-3$, one may take
\begin{equation}
f\equiv\Phi^{(+)}_{\sigma}(X,\xi),\hspace{3mm}f^\star\equiv\Phi^{(-)}_{-\sigma-3}(X,\xi),
\end{equation}  
and substituting these into the Klein-Gordon inner product\footnote{ The Klein-Gordon inner product is defined by
\begin{equation}
\langle f,f'\rangle_{KG}=i\int_\Sigma \Big(f^\star\partial_\mu f'-f'\partial_\mu f^\star\Big)\sqrt{h}n^\mu d^{3}X,
\end{equation}
where $h$ is the determinant of the induced metric on an arbitrary spacelike surface $\Sigma$ and $n^\mu$ is
the forward-directed unit vector normal to $\Sigma$.} yields 
\begin{equation}
\langle f,\,f'\rangle_{KG}= \frac{2^5\pi^4 e^{ \mp i\pi\mu}}{\Gamma(-\sigma)\Gamma(\sigma+3)}\delta(\xi-\xi').
\end{equation}
First we reparametrize the embedding global coordinates as $X:=(\sinh t,\,\cosh t\,\textbf{u})$ and also take the null vector $\xi=(1,\textbf{v})$, where $\textbf{u}$ and $\textbf{v}$ are two unit vectors in the 3-sphere.\footnote{ In the AdS/CFT correspondence, in writing the bulk-to-boundary propagators, similar null vectors have been used (for example, see Ref.\cite{Penedones:2010ue} and its references). The choice of $\xi$, in viewing it as the boundary point of the boundary-to-bulk propagator, is naturally chosen to be at the south pole at future null infinity in de Sitter space.}
Therefore, in terms of $h_\pm$, one obtains
\begin{equation}\label{finalmod}
(X\cdot\xi)^\sigma=\frac{1}{(-\sinh t+\cosh t \,\,\textbf{u}\cdot\textbf{v})^{h_\pm}}.
\end{equation}
If one denotes $\textbf{v}$ and $\textbf{u}$, respectively, by the angle coordinates $\Omega$ and $\Omega'$, then it follows that  
\begin{equation}
\textbf{u}\cdot\textbf{v}= \cos\Theta_3(\Omega,\Omega').
\end{equation}
The quasinormal modes are obtained after evaluating the prime coordinates at the north pole; therefore, the above mode function turns to
\begin{equation}
\Phi^\pm(X,\xi)\sim\frac{1}{(\sinh t+\cosh t \,\,\cos\theta_1)^{h_\pm}},
\end{equation}
where, up to the $i\epsilon$ prescription and the normalization, this mode function can be comparable with the quasinormal mode as stated in (\ref{QNM}). 

A highest-weight quasinormal mode is actually obtained by setting $\Omega=\Omega'=\Omega_{NP}$; this leads to a divergent Klein-Gordon inner product, so that the normalization cannot be done. Instead, the $R$-norm was developed in Ref.\cite{Jafferis:2013qia} for quantizing such modes which have the following properties:
\begin{equation}
\hat{\phi}^\pm(\Omega)=\hat{\phi}^{\pm\dagger}(\Omega_R),\hspace{3mm}\Omega_R(\psi,\theta,\phi)\equiv \Omega(\pi-\psi,\theta,\phi),
\end{equation}
which means the action of $R$ maps an operator defined on the southern hemisphere to one defined on the southern hemisphere of ${\cal I}^+$, and it is defined by  
\begin{equation}
\langle\Phi_1,\Phi_2\rangle_R\equiv \langle\Phi_1,R\Phi_2\rangle_{KG},
\end{equation}
this indeed results in a finite norm for the quasinormal mode which has been discussed in more detail in Ref.\cite{Jafferis:2013qia}.


\section{Generalization to Higher-dimensions and Higher-spin Fields}

The massive free field equation (\ref{mod KG}) in $d+1$ embedding space turns to:
\begin{equation}\label{box}
\Big(\Box_H+\sigma(\sigma+d-1)\Big)(X\cdot\xi)^\sigma=0,
\end{equation}
where the mass term is invariant under the $\sigma\rightarrow -\sigma-(d-1)$, and the box operator is
\begin{equation}
\Box_H=\partial_\alpha\partial^\alpha-d\,\,X_\alpha\partial^\alpha-X_\alpha X_\beta \partial^\alpha\partial^\beta.
\end{equation}
By putting the box operator in (\ref{box}) and after doing some calculation, $\sigma$ is found to be
\begin{equation}
 \sigma(\equiv- h_\pm)=-\frac{d-1}{2}\mp\sqrt{\frac{(d-1)^2}{4}-m^2},
\end{equation}
and its possible values in the complementary series are 
\begin{equation}
-(d-1)<\sigma<0, \hspace{2mm}\mbox{or, equivalently,}\,\, 0<m^2<\frac{(d-1)^2}{4}.
\end{equation}
After some algebra, one obtains the modes in the complementary series as:
\begin{equation}
\Phi^{(\pm)}(X)={\cal N}(\sinh t+\cosh t \cos\theta)^{\sigma},
\end{equation}
where 
\begin{equation}
{\cal N}=\frac{\Gamma(-\sigma)\Gamma(\sigma+d-1)}{2^{d+1}\pi^d }\Big(\theta(X\cdot \xi)e^{\mp i\pi\sigma}+\theta(-X\cdot\xi) e^{\pm i\pi \sigma}\Big).
\end{equation} 

\subsection*{\small{higher-spin generalization}} The method of the previous section can also be used to study higher spins in de Sitter space. The field equation is then given by (see the Appendix)
\begin{equation}
\label{field equation}
\Big({\cal C}^{(s)}_1+\mu^2-\frac{9}{4}+s(s+1)\Big)\psi=0,
\end{equation}
and one needs the relation between the higher-spin Casimir operator and ${\cal C}_1$, nothing that the letter is related to the box operator. For example, in the case of gauge field $s=1$, one has \cite{Gazeau:1999xn}
\begin{equation}
{\cal C}_1^{(1)}K(X)= ({\cal C}_1- 2)K(x) + 2X\partial^T\cdot K - 2\partial X\cdot K,
\end{equation}
where $K(X)$ is a five-vector in embedding space and the dot product is powered by the embedding metric. The four-component vector field $A_\mu(x)$ is locally determined by $K_\alpha(X)$ through the following relations:
\begin{equation}
A_\mu(x)=\frac{\partial X^\alpha}{\partial x^\mu}K_\alpha(X(x)).
\end{equation}
The field equation for a transverse and divergenceless vector field in the complementary series reads as
\begin{equation}
\Big(\Box_H+\mu^2-\frac{9}{4}\Big)K=0, \mbox{or equivalently},\,\,\Big(\Box_H+\sigma(\sigma+3)\Big)K=0,
\end{equation}
where we have used $\sigma=-\frac{3}{2}\mp\mu$. Therefore, the solutions can be found in terms of the introduced plane wave and the polarization function as \cite{Gazeau:1999xn}

\begin{equation}
\begin{array}{l}\label{gauge fie}
K_{1\alpha}(X)=\varepsilon_\alpha^{(\lambda)}(X,\xi)(X\cdot\xi)^\sigma,\hspace*{2mm}\lambda=1,2,3\\
K_{2\alpha}(X)=K_{1\alpha}(X)|_{\sigma\rightarrow-\sigma-3},
\end{array}
\end{equation}
where $\varepsilon_\alpha^{(\lambda)}(X,\xi)$ are the three polarization vectors defined by
\begin{equation}
\varepsilon_\alpha^{(\lambda)}(X,\xi)=(\sigma+3)\Big(Z^{T(\lambda)}_\alpha-(\sigma+2)\frac{Z^{(\lambda)}\cdot X}{X\cdot\xi}\xi^T_\alpha\Big).
\end{equation}
We would like to mention that the modes (\ref{gauge fie}) in the principal series cover the gauge field of the Minkowski space in the flat limit.
 
After evaluating on the north pole similar to the case of the scalar field and in global coordinates, the modes read as
\begin{equation}
K^\pm_{\alpha}(X)\sim\varepsilon_\alpha^{(\lambda)}(X,\xi)\frac{1}{(\sinh t+\cosh t \cos\theta)^{h_\pm}}.
\end{equation}
Note that there appears a constant five-vector in obtaining the polarization vector. It is actually fixed by demanding that, at the zero-curvature limit $(H\rightarrow0),$  the polarization vector must tend to the three polarization vector in Minkowski space, $\varepsilon_\mu^{(\lambda)}(k)$, and one obtains 
\begin{equation}
Z_\alpha^{(\lambda)}=\Big(\varepsilon_\mu^{(\lambda)}(k),Z_4^{(\lambda)}=0\Big).
\end{equation}  

Let us repeat the calculations for the spinor field\footnote{Dirac quasinormal modes of $d$-dimensional de Sitter space have been considered in Ref.\cite{LopezOrtega:2012vi} by finding the exact solution of the Dirac equation in de Sitter space.} $s=\frac{1}{2}$, where for a spinor field $\psi$ one has
\begin{equation}\label{spin1/2field eq}
\Big({\cal C}_1^{(1/2)}+\mu^2-\frac{3}{2}\Big)\psi=0.
\end{equation}
The second-order Casimir operator for spin $\frac{1}{2}$ can be
written as 
\begin{eqnarray}
{\cal C}_1^{(1/2)}&=&-\frac{1}{2}M_{\alpha\beta}M^{\alpha\beta}-\frac{1}{2}S_{\alpha\beta}S^{\alpha\beta}-S_{\alpha\beta}M^{\alpha\beta}\nonumber\\
&=&{\cal C}_1-\frac{5}{2}+\not X\not\partial^T,
\end{eqnarray}
where $S_{\alpha\beta}$ is the spin part of the Casimir operator which is given by (\ref{s operator}) and $ \gamma_\alpha X^\alpha\equiv \not X$. Inserting the latter in (\ref{spin1/2field eq}) leads to
\begin{equation}\label{ref}
\Big(\Box_H+\not X\not\partial^T+\mu^2-4\Big)\psi=0,
\end{equation} 
where ${\cal C}_1=\Box_H$. This second-order equation should be solved to obtain the fields.\footnote{The solutions of
Eq.(\ref{ref}) in the principal series are obtained in Ref.\cite{Bartesaghi:2001ea} as the spinor fields in de Sitter space,
and in the flat limit these solutions tend to the Dirac spinors} Let us introduce ${\cal D}\equiv-\not X\not\partial^T+2$; then the above equation turns to
\begin{equation}
({\cal D}+\mu)({\cal D}-\mu)\psi=0.
\end{equation}
The solution can be written in terms of a scalar plane wave and two polarization spinors as
\begin{eqnarray}
\psi_1&=& u_+^{(\lambda)}(\xi)(X\cdot \xi)^\sigma\\ \nonumber
\psi_2&=& u_-^{(\lambda)}(\xi)\not X\not\xi(X\cdot \xi)^\sigma,
\end{eqnarray}
where $\lambda=1,2$ and $u_{\pm}(\xi)$ are two polarization spinors. Note that the homogeneity consideration of (\ref{ref}) reveals that $\sigma= -2-\mu$. The introduced spinors are the same as the principal series; one may do the calculations first in the rest frame and then by acting the group generators the generic one can be achieved. Let us set $\xi^\alpha=(k^\mu,\pm1)$ and the rest frame is given by $\underline{\xi}=(\underline{k},\pm1)$, where $\underline{k}\equiv(1,\overrightarrow{0})$; the spinors are obtained as
 \begin{eqnarray}
 u_\pm^{(\lambda)}(k)=\frac{\pm \gamma_\mu k^\mu\,\gamma^4+1}{\sqrt{2(1+k^0)}}u_\pm^{(\lambda)}(\underline{k})\\\nonumber
 u_\pm^{(\lambda)}(\underline{k})=\frac{1}{\sqrt{2}}\left( \begin{array}{clcr} \chi^\lambda \\ \pm\chi^\lambda \\
    \end{array} \right)
 \end{eqnarray}
 where $\chi^1=\left( \begin{array}{clcr} 1 \\ 0 \\
    \end{array} \right)$ and $\chi^2=\left( \begin{array}{clcr} 0 \\ 1 \\
    \end{array} \right)$.
Thus the globally defined spinor fields in de Sitter space when they are calculated at the north pole are found as:
\begin{equation}
\psi_\pm^\lambda\sim u_\pm^{(\lambda)}(k)\frac{1}{(\sinh t+\cosh t \cos\theta)^{2+\mu}}.
\end{equation}

\section{Conclusions}

In the context of dS$_4$/CFT$_3$ correspondence, for every spin-zero primary CFT$_3$ operator ${\cal O}$ of conformal weight $h$, there is a bulk scalar field $\Phi(X)$ with mass $m^2=h(3-h)$. Boundary correlators of ${\cal O}$ are then related by a rescaling to bulk $\Phi(X)$ correlators whose arguments are pushed to the boundary at ${\cal I}^+$.  As in AdS/CFT, the quasinormal modes are just the analogues of the bulk-to-boundary propagator in which is obtained by taking one of the arguments of the bulk-to-bulk propagator to the boundary. In the embedding space extension, the bulk-to-boundary propagator is just proportional to $\frac{1}{({\cal P}\cdot X)^h}$, where $h$ is the conformal dimension of the dual operator, ${\cal P}$ is a null vector in the embedding space and $X$ is a point on AdS. This formalism has been discussed in several papers (see \cite{Penedones:2010ue} and references therein). Here, we extended the embedding formalism to de Sitter in more detail. Explicitly, in this work we showed that the quasinormal modes in de Sitter space are related to one class of the UIR of the de Sitter group, namely, the complementary series, when the so-called plane waves are utilized. From the group theoretical point of view and after making use of the plane waves which are defined in the embedding space, we reproduced the scalar QNMs in de Sitter space. This approach may be generalized to higher-spin and dimensions in a simple way, as we did for finding the gauge and spinor quasinormal modes. The QNMs for various spins sometime are obtained through some seemingly different details of computations. So, this method provides a straightforward and unified way of deriving them. Furthermore, knowing the expressions for all higher spin might prove to be useful in the context of higher-spin dS$_4$/CFT$_3$.

Moreover, in Refs.\cite{Ng:2012xp, Jafferis:2013qia}, the QNMs have been considered only for the light massive scalar fields $0<m^2<\frac{9}{4}$; as mentioned this mass range coincides with the complementary series of the UIR of the de Sitter group. However, group theoretical consideration allows us to generalize the discussion to massive fields as long as $m^2>\frac{9}{4}$, leads to imaginary $\mu$. In the Appendix we have reviewed the UIRs to show that, by setting $\mu\rightarrow i\mu$, the principal series of the UIR is achieved. In the principal series, $\sigma$ (and hence the weights $h$) has an imaginary part, and the corresponding modes, while still being damped near the boundaries, have an oscillatory behavior in the bulk. In addition, in the flat limit, the corresponding modes in this series reduce to their counterpart in flat space.

On the other hand, these nonoscillatory modes have imaginary frequencies because of their damping characteristic; therefore, associating a Hermitian operator to them is somewhat problematic. Group theoretical consideration, however, relates them to the modes associated with the Casimir operator of the de Sitter group. The damping characteristic may lead to the infrared behavior of such fields. As mentioned, in the mass range this series corresponds to the light masses, thus one should note such a bad behavior of particles at the early Universe when all the particles of the standard model have light masses.

{\subsection*{Acknowledgements}

Special thanks to M. Alishahiha for first suggesting the idea and also for his useful comments and hints. It has been a great pleasure discussing with M.V. Takook t hede Sitter plane wave method. I also thank M. Reza Mohammadi-Mozafar and A. Faraji for their comments. This work was supported in part by Islamic Azad
University Central Tehran Branch.

\section*{Appendix}
The isometry group of de Sitter is the ten-parameter homogeneous Lorentz group $SO_0(1,4)$, and the realization of the Lie algebra of the dS group can be understood by the ten Killing vectors:
\begin{equation}
K_{\alpha\beta}=X_\alpha\partial_\beta-X_\beta\partial_\alpha.
\end{equation}
There are two coordinate invariant Casimir operators, and the fields can be classified according to the irreducible representations of the de Sitter group which are labeled by the eigenvalues
associated with the Casimir operators. The two Casimir operators are defined by
\begin{equation}
\begin{array}{l}
{\cal C}_1=-\frac{1}{2}L_{\alpha\beta}L^{\alpha\beta},\\
{\cal C}_2=-W_{\alpha}W^{\alpha},\,\,\,\,W_{\alpha}=-\frac{1}{8}\epsilon_{\alpha\beta\gamma\sigma\eta}L^{\beta\gamma}L^{\sigma\eta},
\end{array}
 \end{equation}
where $\epsilon_{\alpha\beta\gamma\sigma\eta}$ is the usual antisymmetric tensor. $L_{\alpha\beta}=M_{\alpha\beta}+S_{\alpha\beta}$, in which the orbital part is defined by $M_{\alpha\beta}=-i(x_{\alpha}\partial_{\beta}-x_{\beta}\partial_{\alpha})$ while the spinorial part $S_{\alpha\beta}$ acts
on the indices of a given function in a certain way; for example, in the case of rank 2 tensor field one has:
$$ S_{\alpha\beta}{\cal K}_{\gamma\delta}=-i(\eta_{\alpha\gamma}{\cal K}_{\beta\delta}-\eta_{\beta\gamma} {\cal K}_{\alpha\delta} +
 \eta_{\alpha\delta}{\cal K}_{\beta\gamma}-\eta_{\beta\delta}
 {\cal K}_{\alpha\gamma}),$$
and in the spinorial case, we have \cite{Bartesaghi:2001ea} 
\begin{equation}\label{s operator}
S^{(1/2)}_{\alpha\beta}=\frac{-i}{4}[\gamma_\alpha,\gamma_\beta].
\end{equation}
Note that here one needs five $4\times4$ $\gamma_\alpha$  matrices which are the generators of the Clifford algebra based on the metric $\eta_{\alpha\beta}$, namely, 

\begin{equation}
\gamma^\alpha\gamma^\beta+\gamma^\beta\gamma^\alpha=2\eta^{\alpha\beta},\,\,\,\,\gamma^{\alpha\dagger}=\gamma^0\gamma^\alpha\gamma^0 .\nonumber
\end{equation}
Explicitly, the $\gamma$ matrices used in this paper are
$$ \gamma^0=\left( \begin{array}{clcr} I & \;\;0 \\ 0 &-I \\ \end{array} \right)
      ,\gamma^4=\left( \begin{array}{clcr} 0 & I \\ -I &0 \\ \end{array} \right) , $$ \begin{equation}
   \gamma^1=\left( \begin{array}{clcr} 0 & i\sigma^1 \\ i\sigma^1 &0 \\
    \end{array} \right)
   ,\gamma^2=\left( \begin{array}{clcr} 0 & -i\sigma^2 \\ -i\sigma^2 &0 \\
      \end{array} \right)
   , \gamma^3=\left( \begin{array}{clcr} 0 & i\sigma^3 \\ i\sigma^3 &0 \\
      \end{array} \right),
      \end{equation}
where $\sigma_i$ are the Pauli matrices and $I$ is a $2\times2$ unit
matrix. In the case of spinor fields, $s=l+\frac{1}{2}$, one can write: $S^{(s)}_{\alpha\beta}=S_{\alpha\beta}+S^{(1/2)}_{\alpha\beta}$.
  
In fact, only ${\cal C}_1$ is relevant in characterizing fields since it leads to quadratic field equations, note that the second one automatically vanishes for spin-zero fields. Following Dixmier \cite{Dix}, one gets a classification scheme using
a pair $(p,q)$ of parameters involved in the following possible spectral values of the Casimir operators:
\begin{equation}
\langle{\cal C}_1\rangle=\left(-p(p+1)-(q+1)(q-2)\right),\qquad\quad
\langle{\cal C}_2\rangle=\left(-p(p+1)q(q-1)\right).
\end{equation}
Two introduced parameters $p$ and $q$ could potentially play the role of spin and mass similar to flat space where mass and spin are the eigenvalues of the Poincar\'e group that label the UIRs. Accordingly, there are three types of unitary irreducible representations. 
\subsubsection*{Principal series $U_{s,\nu}$ }
This series is also called `massive' representations, and the UIRs are labeled by $(p,q)=(p,\tfrac{1}{2} \pm i\nu)$, where $i\nu\equiv\mu=\sqrt{\frac{9}{4}-m^2}$. The meaning of spin as carried out by $p$ and $q$ is related to the mass via the dimensionless parameter $\mu$. In the language of the mass term \cite{Bros:2006gs, Leblond:2002tf}, the principal series is defined by the inequality $m^2>\frac{9}{4}$  which leads $\sigma$ and hence the weight $h_\pm$ to have an imaginary part [see relation \ref{h}], and the corresponding modes oscillate in the bulk. It is proved that in the flat limit these UIRs reduce to the corresponding wave equations in Minkowski space. The eigenvalue of the Casimir operator is fixed as
\begin{equation}
\langle{\cal C}_1\rangle = \big(\tfrac{9}{4} - \mu^2 -s(s+1) \big).
\end{equation}
 One has to distinguish between
\bei\itemsep=0pt
\item[$(i)$] the integer spin principal series: $\nu \in \R$,  $s=1,2, \dots$, 
\item[$(ii)$] and the half-integer spin principal series: $\nu \neq 0$, $s= \frac{1}{2}, \frac{3}{2}, \frac{5}{2}, \dots$.
\eni
In both cases, $U_{s,\nu}$ and $U_{s,-\nu}$ are equivalent.

\subsubsection*{Complementary series $V_{s,\mu}$}
Similar to the principal series $p=s$ has a spin meaning and $(p,q)=(p,\tfrac{1}{2} \pm \mu)$. For the complementary series,
the effective mass falls in the range $0 < m^2 < \frac{9}{4}$. The corresponding modes are nonoscillatory asymptotically, because $\sigma$ becomes a real quantity. And one obtains
\begin{equation}
\langle{\cal C}_1\rangle  =  \big(\tfrac{9}{4} - \mu^2 -s(s+1)\big).
\end{equation}
We have to distinguish between
\bei\itemsep=0pt
\item[$(i)$] {\it the scalar case} $V_{0,\mu}$, $\mu \in \R$, $0 < \vert \mu \vert < \frac{3}{2}$, and
\item[$(ii)$] {\it the spinorial case} $V_{s,\mu}$,  $0 < \vert \mu \vert < \frac{1}{2}$, $s = 1, 2, 3, \dots$.
\eni
In both cases, $V_{s,\mu}$ and $V_{s,-\mu}$ are equivalent.

\subsubsection*{Discrete series $\Pi^{\pm}_{p,q}$}

Parameter $q$ has a spin meaning, and this series corresponds to $m^2<0$. This tachyon field when approaching ${\cal I}^\pm$ grows without bound. Similarly one has to distinguish between
\bei\itemsep=0pt
\item[$(i)$]{\it the scalar case}
$\Pi_{p,0}$, $p=1,2, \dots$. These representations lie at the ``lowest limit'' of the discrete series and are not square integrable,
\item[$(ii)$] {\it the spinorial case} $\Pi^{\pm}_{p,q}$, $q>0$,
$p= \frac{1}{2}, 1, \frac{3}{2}, 2, \dots$, $q=p, p-1, \dots, 1$ or $\frac{1}{2}$ for $q = \frac{1}{2}$, the representations
$\Pi^{\pm}_{p,\frac{1}{2}}$ are not square integrable.
\eni

In any case, the second-order wave equation can be written as
\begin{equation}
({\cal C}^{(s)}_1-\langle{\cal C}^{(s)}_1\rangle)\psi=0,\hspace*{2mm}\mbox{where}\,\,\, \langle{\cal C}^{(s)}_1\rangle =  \big(\frac{9}{4}-\mu^2 -s(s+1)\big).
\end{equation}

\end{document}